\documentclass[12pt]{article}
\usepackage{afterpage,amsfonts,amsmath,amssymb,array}
\usepackage{subfigure, graphicx}
\usepackage{fancyhdr}
\usepackage[blackqed,whiteqed]{nccthm}

\def\cd desktop/General_sty

\usepackage{\dir chicago}

\renewcommand{\theequation}{\mbox{\arabic{section}.\arabic{equation}}}

\begin{document}
\thispagestyle{empty}

{\bf  Exact Pricing and Hedging Formulas of Long Dated Variance Swaps under a $3/2$ Volatility Model}   \ \\

Leunglung Chan, School of Finance and Economics, University of
Technology, Sydney, PO Box 123, Broadway, NSW 2007, Australia\ \\
(Leunglung.chan@uts.edu.au)\ \\

Eckhard Platen, School of Finance and Economics and Department of
Mathematical Sciences, University of Technology, Sydney, PO Box 123,
Broadway, NSW 2007, Australia (Eckhard.Platen@uts.edu.au)
 \ \\
\line(1,0){385} \ \\

{\bf Abstract} 

This paper investigates the pricing and hedging of variance swaps under a $3/2$ volatility model. Explicit pricing and hedging formulas of variance swaps are obtained under the benchmark approach, which only requires the existence of the num\'{e}raire portfolio. The growth optimal portfolio is the num\'{e}raire portfolio and used as num\'{e}raire together with the real world probability measure as pricing measure. This pricing concept provides minimal prices for variance swaps even when an equivalent risk neutral probability measure does not exist.

{\em 1991 Mathematics Subject Classification}: Primary 62P05; secondary 60G35, 62P20\\
{\em JEL Classification}: G10, G13\\
{\em Key words}: $3/2$ volatility model, variance swap, num\'{e}raire
portfolio, squared Bessel process, confluent hypergeometric functions\\
{\bf{Date: DEC. 10, 2010}}

 \ \\
\line(1,0){385}
\section{Introduction}
 \label{sec.1}

  A significant source of risk in a financial market is the uncertainty of the volatility of equity indices. During  financial turmoil, volatility risk is of extreme importance to investors and derivative traders. Additionally, due to large and frequent shifts in the volatility of various assets in volatile periods, there is a growing practical need for appropriate models that allow for the realistic modeling of volatility, the pricing of related financial instruments and the hedging volatility risk. In 1993, the Chicago Board Options Exchange (CBOE) introduced a volatility index, the VIX, see Figure $1.1$, based on the implied volatilities of options on the
S\&P 500 index, see Figure $1.2$. The volatility of this diversified index has attracted the most attention in the current literature on volatility derivatives. As such, this paper singles out as the security of interest a similarly diversified equity index and studies its volatility and related derivatives.

\begin{figure}[ht]
\centering
{\includegraphics[width=7.2cm]{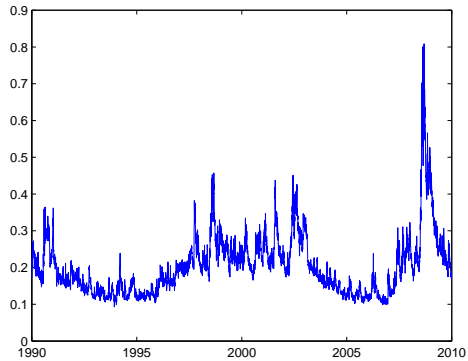}}
\caption{VIX from $1990$ to $2010$.}
\end{figure}
\begin{figure}[ht]
\centering
{\includegraphics[width=7.2cm]{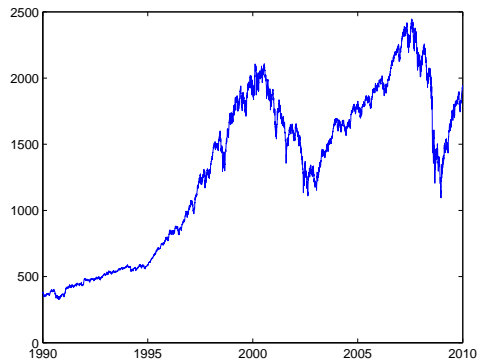}}
\caption{S\&P $500$ from $1990$ to $2010$.}
\end{figure}

Variance swaps have been
 traded in over-the-counter markets since the collapse of Long Term Capital Management in late 1998. In particular,
variance swaps on stock indices have been traded actively as a hedge for volatility risk.
Investors and fund managers alike, have developed an interest in volatility derivatives since these instruments may substantially increase the value of their holdings, even if the equity market index experiences a major drawdown. At least theoretically, these derivatives can  provide some protection against severe market downturns. How effective such portfolio insurance is, from a macro-economic view point, remains an open question. In particular, when a large and increasing number of pension funds, insurance companies and other investors rely on this type of insurance, it is not clear whether the sellers of variance swaps will be able to serve their obligations in a market crash.

There exists a substantial literature on volatility modeling, and one may refer to Cont and Tankov (2004) as one of  many references.
Several papers, which explore stochastic volatility models, have pointed at the seemingly undesirable property of some models, where the moments of squared volatility of higher order than one may become infinite in finite time. Examples are given in Andersen and Piterbarg (2007), Lion and Musiela (2007) and Glasserman and Kim (2010).
Furthermore, there exist various papers discussing the general problem of pricing and hedging variance swaps, including Brenner et al. (1993), Gr\"{u}nbuchler and Longstaff (1996), Carr and Madan (1999), Chriss and Morokoff (1999), Demeterfi et al. (1999), Brockhaus and Long (2000), Matytsin (2000), Javaheri et al. (2002), Swishchuk (2004), Howison et al. (2004), Carr et al. (2005), Windcliff et al. (2006), Zhang and Zhu (2006), Zhu and Zhang (2007), Sepp (2007),  Elliott, Siu and Chan (2007) and Carr and Lee (2009).

Recently, Carr and Sun (2007), as well as, Itkin and Carr (2009) discussed in two interesting papers the pricing of variance swaps under the, so called, $3/2$ volatility model. Earlier versions of $3/2$ volatility models were studied, for instance, in Cox et al. (1980, 1985), Platen (1997), Ahn and Gao (1999), Heston (1999), Lewis (2000), Andreasen (2001), Platen (2001) and Spencer (2003).

The current paper studies the case of a model which it calls the real world $3/2$ volatility model.
Explicit pricing formulas of variance swaps are obtained. This model generalizes the minimal market model (MMM), which was introduced by Platen (2001) and does not have an equivalent risk neutral probability measure. The real world pricing concept of the benchmark approach aims to identify the minimal prices for variance swaps.

The paper is organized as follows: Section $2$ introduces variance swaps, and relates these to the popular log-contract hedging. Section $3$ introduces a diversified equity index model. Section $4$ demonstrates the estimation of parameters. Section $5$ studies the risk neutral $3/2$ volatility model. Section $6$ considers the real world $3/2$ volatility model. Section $7$ introduces the real world pricing concept. The analytical pricing formulas for variance swaps are obtained in Section $8$. Section $9$ discusses hedging issues. Section $10$ shows some numerical results. Finally, Section $11$ concludes.

\section{Variance Swaps}
\renewcommand{\theequation}{2.\arabic{equation}}
\setcounter{equation}{0}
For simplicity, assume throughout the paper that one has zero interest rate. A variance swap is a forward contract on
annualized variance, which is the square of the realized annualized volatility.
Let $\sigma^2_{0,T}$ denote the realized annualized variance of the log-returns of a diversified equity index or related futures over the
life of the contract, such that
\begin{eqnarray}
\sigma^2_{0,T}:= \frac {1} {T} \int^{T}_{0} \sigma^{2}_{u} d u \ ,
\end{eqnarray}
where $\sigma_{u}$ denotes the volatility of the index or futures at time $t$.
Assume that one can trade the futures or index at discrete times $t_{i}=i\Delta$ for $i\in\{0,1,...\}$ with time step size $\Delta>0$. The period $\Delta$ between two successive potential trading times is typically the length of one day. If $F_{t_{i}}$ denotes the discounted index price at time $t_{i}$, then the return $G_{t_{i}}$ for the period before this time is defined as
 \begin{eqnarray}
 G_{t_{i}}=\frac{F_{t_{i}}-F_{t_{i-1}}}{F_{t_{i-1}}}
 \end{eqnarray}
for $i\in\{1,2,...\}$.
In practice, variance swaps are often based on the realized variance,
evaluated from daily closing prices, of the form
\begin{eqnarray}
{\hat{\sigma}}^{2}_{0,t_{n}}:=\frac{N}{n}\sum_{i=1}^{n}\bigg(G_{t_{i}}\bigg)^{2}\approx \frac{N}{n}\sum_{i=1}^{n}\bigg\{\ln\bigg(\frac{F_{t_{i}}}{F_{t_{i-1}}}\bigg)\bigg\}^{2}:={\tilde{\sigma}}^{2}_{0,t_{n}}, \end{eqnarray}
where $T=t_{n}=\sum_{i=1}^{n}(t_{i}-t_{i-1})=\frac{n}{N}$, and $N$ is the number of trading days per year. Hence, variance swaps with idealized
payoffs, depending on the realized variance, as defined in (2.1),
are only approximations to those of the actual contracts. As indicated in (2.3), both returns or log-returns may be used to define payoffs on realized variance. Furthermore, by (2.1) and (2.3) it becomes clear that the basic elements of this type of payoffs are some form of squared volatility to be paid at a given time. This is also why the paper will focus on the pricing of a payoff that delivers squared volatility.

 As is common in most of the literature, $(\Omega, {\cal A}_{T},\underline{\cal{A}}, Q)$ denotes some assumed risk neutral, filtered probability space.  Here $Q$ is denoting the assumed risk neutral probability measure. The filtration $\underline{\cal{A}}=({\cal{A}}_t)_{t\in[0,T]}$ models the evolution of the market information over time, where ${\cal A}_t$ describes the information available at time $t\in [0,T]$.
Let $K_v$ denote the delivery price for realized variance and $L$ the notional
amount of the swap in dollars per annualized variance point.
Then, the payoff of the variance swap at expiration time $t_{n}$ is given
by $L ({\tilde{\sigma}}^{2}_{0,t_{n}} - K_v )$. Intuitively, the buyer of the variance
swap will receive $L$ dollars for each point by which the realized
annual variance ${\tilde{\sigma}}^{2}_{0,t_{n}}$ has exceeded $K_v$.  Under the risk neutral approach, the value of the variance swap can be
evaluated as the expectation of its discounted payoff with respect to the
assumed risk neutral measure $Q$. This value is equivalent to the value
of a forward contract on future realized variance with strike price $K_v$.

The risk neutral value $\hat{V}$ of the variance swap at time $t=0$ over the period $[0, t_{n}]$ is given by the expression
\begin{eqnarray}
{\hat{V}} &=& E^{Q} \bigg(L ({\tilde{\sigma}}^{2}_{0,t_{n}} - K_v )\bigg) \nonumber\\
&=& L E^{Q} ({\tilde{\sigma}}^{2}_{0,t_{n}}) - L K_v   \ ,
\end{eqnarray}
where $E^{Q}$ denotes expectation under the assumed risk neutral measure $Q$.
Hence, the valuation of the variance swap relies on calculating the risk neutral expectation $E^{Q} ({\tilde{\sigma}}^{2}_{0,t_{n}})$ of the realized variance. Note by (2.3) and (2.1) that this involves the computation of a sum of risk neutral expectations of squared volatility. Therefore, the problem can be reduced to evaluating the risk neutral expectation of squared volatility, as mentioned earlier.

To highlight some of the critical issues with the current methodology, recall the popular log-contract hedging approach to realized annualized variance, as employed for instance in Neuberger (1990), Dupire (1993), Demeterfi et al. (1999) and Carr and Lee (2009). In the above literature the payoff ${\tilde{\sigma}}^{2}_{0,t_{n}}$, appearing in (2.3), is usually approximated by the expansion
 \begin{eqnarray}
{\tilde{\sigma}}^{2}_{0,t_{n}}&=&\sum_{i=1}^{n}\frac{2N}{n}\bigg(\frac{1}{F_{t_{i-1}}}-\frac{1}{F_{t_{0}}}\bigg)(F_{t_{i}}-F_{t_{i-1}})+\bigg[\int_{0}^{F_{t_{0}}}\frac{2N}{nK^{2}}(K-F_{t_{n}})^{+}dK\nonumber\\
&+&\int_{F_{t_{0}}}^{\infty}\frac{2N}{nK^{2}}(F_{t_{n}}-K)^{+}dK\bigg]-\frac{N}{3n}\sum_{i=1}^{n}G_{t_{i}}^{3}+\sum_{i=1}^{n}\mathcal{O}(G_{t_{i}}^{4}),
\end{eqnarray}
where the return $G_{t_{i}}<<1$ is assumed to be ``small''. To be precise in our notation, $\mathcal{O}(G_{t_{i}}^{q})$ represents a term that is smaller in absolute value than some constant times $|G_{t_{i}}|^{q}$ when $G_{t_{i}}$ asymptotically vanishes.
The first two terms on the right-hand side of (2.5) form the profit and loss from a dynamic position in the index and a static position in options on the index, respectively. For the dynamic component, one holds $\frac{2N}{n}\bigg(\frac{1}{F_{t_{i-1}}}-\frac{1}{F_{t_{0}}}\bigg)$ futures contracts on the index from day $t_{i-1}$ to day $t_{i}$. For the static component, one holds $\frac{2N}{nK^{2}}dK$ European put options with strikes $K$ below the initial index value $F_{t_{0}}$. One also holds $\frac{2N}{nK^{2}}dK$ European call options having strikes greater than the initial index value $F_{t_{0}}$. As such, one holds more puts than calls. In the literature it is argued that the third and fourth terms on the right hand side of (2.5) represent the most significant source of error when approximating  the payoff ${\tilde{\sigma}}^{2}_{0,t_{n}}$. In particular, the fourth power of the return appears in the remainder term in (2.5). This paper emphasizes that the expansion (2.5) is of a pathwise nature and relies on the assumption that $|G_{t_{i}}|$ is ``sufficiently small'' for all scenarios, which is a rather strong and potentially unrealistic assumption, as we will discuss below.

 It was shown in Platen and Rendek  (2008), that the Student-$t$ distribution with approximately four degrees of freedom is the typical estimated log-return distribution of diversified world stock indices when such indices are denominated in a currency. Similar empirical evidence is provided in Markowitz and  Usmen (1996a, 1996b) for the S\&P 500. However, it is well-known that the fourth moment of the Student-$t$ distribution with four degrees of freedom is infinite. In view of the assumed representation (2.5) this raises the critical question whether the error truncation in (2.5) is a sensible one for practical purposes. It will turn out that the real world $3/2$ volatility model generates Student-$t$ distributed log-returns for the underlying equity index. In the case of the stylized minimal market model (MMM), these log-returns have four degrees of freedom. But also for generalizations of the MMM with higher degrees of freedom it is a matter of fact that extreme log-returns of diversified equity indices occur from time to time, as they arise also in reality rarely but also with some nonvanishing likelihood. Under the $3/2$ volatility model this happens with a probability that is not negligible. Therefore, one may question whether it is prudent to assume in (2.5) that $|G_{t_{i}}|$ is always ``sufficiently small''.

Therefore, when pricing volatility derivatives, one has to deal with expectations.
Hence, the expectation of the final term of (2.5) may create a problem in this respect since it can be extremely large or even infinite under models that generate Student-$t$ log-returns with realistic degrees of freedom. Therefore, based on these facts extreme caution should be taken when pricing and hedging volatility derivatives using the  type of expansion given in (2.5). The authors of the current paper would like to argue that the reliance on the assumption that the expectation of the error term in (2.5) is ``small'' should be avoided. Alternative pricing and hedging methods should be employed that do not use any such assumption. To visualize this problem, Figure (2.3) exhibits the cumulative fourth term on the right hand side of (2.5) for S\&P $500$ index daily log-return data, from 1990 until 2010.
 One notes that during the period from September to October 2008 the cumulative fourth term increased by approximately $300\%$ in just a few days. This means that $G_{t_{i}}^{4}$ itself must have experienced extra ordinarily high values. This figure indicates, at least visually, that the fourth term in (2.5) cannot be easily neglected as a remainder term that is always ``sufficiently small''. In reality, a severe approximation error has to be expected when employing the widely propagated methodology, which uses a truncated version of the expansion (2.5).

\begin{figure}[ht]
\centering
{\includegraphics[width=7.2cm]{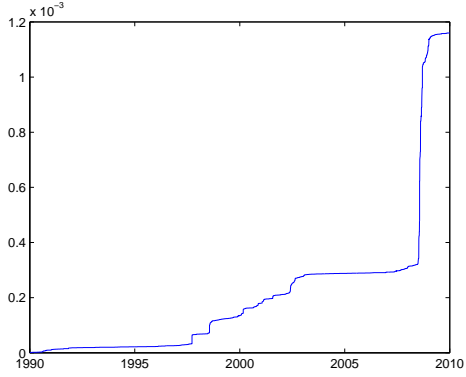}}
\caption{The cumulative fourth term, $\sum_{i=1}^{n}\mathcal{O}(G_{t_{i}}^{4})$, using daily S\&P $500$ data from $1990$ to $2010$.}
\end{figure}

To avoid these problems in the valuation of variance swaps, we suggest to adopt the benchmark approach, see Platen and Heath (2006). In fact, as described in Section $7$, the variance swap will be
evaluated under this approach by using the expectation of its benchmark payoff with respect to the real world probability measure $P$. In principle, this is the value
of a forward contract on the future realized variance ${\tilde{\sigma}}^{2}_{0,T}$ with strike price $K_v$.
Since one considers all quantities under the real world probability measure, this excludes the modeling error that potentially arises from the assumption on the existence of an equivalent risk neutral probability measure.

\section{A Model for a Diversified Equity Index}
\renewcommand{\theequation}{3.\arabic{equation}}
\setcounter{equation}{0}
Denote by $r_t$ the predictable interest rate at time $t$.
The savings account value at time $t$ is denoted by $S_{t}^{0}$ and given by the SDE
\begin{eqnarray*}
\frac{dS_t^{0}}{S_{t}^{0}}=r_{t} dt,
\end{eqnarray*}
where $r_t$ may be a stochastic or deterministic interest rate.

Denote by $F_{t}$ the value of a discounted diversified equity index at time $t\ge 0$, and assume that
\begin{eqnarray}
F_{t}=A_{t}Y_{t}
\end{eqnarray}
for $t\ge 0.$ Here $A_{t}$ stands for the average growth of the index, where
\begin{eqnarray}
A_{t}=\tilde{\alpha}\exp\{\eta t\}
\end{eqnarray}
for $t\ge 0$, with $\tilde{\alpha}>0$ and long term net growth rate $\eta\in \Re$.
The mean reverting dynamics of the index around its average exponential increase is modeled by an ergodic process $Y=\{Y_{t}, t\ge 0\}$, which is assumed to satisfy the SDE
\begin{eqnarray}
dY_{t}=(a_{1}-b Y_{t})dt+\sqrt{2\gamma Y_{t}}dW(t)
\end{eqnarray}
for $t\ge 0$ with $Y_{0}>0$. Here $b>0$ is the speed of adjustment, which results for $Y_{t}$ in the reference level $\frac{a_{1}}{b}$ as long run mean. The process $Y$ is a square root process of dimension $\delta=\frac{2a_{1}}{\gamma}$, where we assume $\delta>2.$ Furthermore, $W=\{W(t),t\ge 0\}$ is a standard Wiener process.

By an application of the It\^{o} formula one obtains from (3.1), (3.2) and (3.3) for $F_{t}$ the SDE
\begin{eqnarray}
dF_{t}=F_{t}(\mu_t dt+\sigma_{t}dW_{t})
\end{eqnarray}
for $t\ge 0$ with $F_{0}=A_{0}Y_{0}$. Here one has the expected rate of return
\begin{eqnarray}
\mu_{t}=\eta-b+\frac{a_{1}}{Y_{t}}
\end{eqnarray}
and the volatility
\begin{eqnarray}
\sigma_{t}=\sqrt{\frac{2\gamma}{Y_{t}}}.
\end{eqnarray}
Therefore, the squared volatility $v_{t}=\sigma_{t}^{2}$ satisfies under this model the SDE
\begin{eqnarray}
dv_{t}&=&d\bigg(\frac{2\gamma}{Y_{t}}\bigg)\nonumber\\
&=& \bigg(b v_{t}+(1-\frac{a_{1}}{2\gamma})v_{t}^{2}\bigg)dt - v_{t}^{3/2} dW_{t}.\nonumber\\
\end{eqnarray}
for $t\ge 0$.
The dynamics for the volatility of the index turn out to be those of a $3/2$ volatility model, see Cox et al. (1980, 1985), Platen (1997), Ahn and Gao (1999), Heston (1999), Lewis (2000), Andreasen (2001), Platen (2001), Spencer (2003), Carr and Sun (2007) and Itkin and Carr (2009). Furthermore, they are ergodic, and, thus, due to their stationary density remain always within a certain range, even over long periods of time.

From (3.4) it follows that the market price of risk $\theta_{t}$ is under the above model of the form
\begin{eqnarray}
\theta_{t}=\frac{\mu_{t}}{\sigma_{t}}=\bigg(\eta-b+\frac{a_{1}}{Y_{t}}\bigg)\sqrt{\frac{Y_{t}}{2\gamma}}=\bigg( (\eta-b)\sqrt{Y_{t}}+\frac{a_{1}}{\sqrt{Y_{t}}} \bigg)\frac{1}{\sqrt{2\gamma}}.\nonumber\\
\end{eqnarray}
This shows that the market price of risk becomes large when the market index attains relatively small values. As one can see from Figure $1.1$ and Figure $1.2$, this reflects well reality.

\section{Parameters Estimation}
\renewcommand{\theequation}{4.\arabic{equation}}
\setcounter{equation}{0}
This section shows the estimation of parameters of the model. A set of parameters which are needed to be estimated are $\tilde{\alpha}$, $\eta$, $a_{1}$, $b$ and $\gamma$. Figure $4.4$ plots the logarithm of the S\&P $500$ total return index, observed in US dollars from 1920 until March 2010.  Furthermore, it also plots a linearly regressed function of time: $0.1028t-1.3929$.
By a least square fit of the logarithm of the S\&P500, we find $\tilde{\alpha}=0.248$ and $\eta=0.1028$.
\begin{figure}[ht]
\centering
{\includegraphics[width=7.2cm]{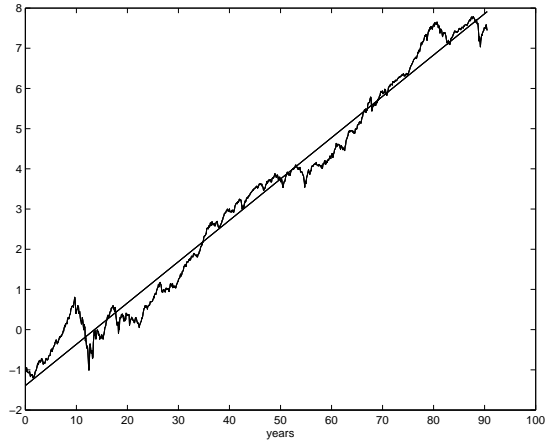}}
\caption{Logarithms of the S\&P $500$ from $1920$ to $2010$ as well as linear fit.}
\end{figure}

 Remarkably, this model reduces to the stylized MMM proposed in Platen (2001) when setting parameters $a_{1}=1$, $b=\eta$ and $\gamma=\frac{1}{2}$, see also Platen and Heath (2006).
In this special case the model has only two parameters, $\tilde{\alpha}$ and $\eta$.
One notes its long term growth with seemingly stationary fluctuations around the average growth.
By taking the logarithm on both sides of (3.1) it follows that
\begin{eqnarray}
\ln(F_{t})=\eta t+\ln(Y_{t})+\ln(\tilde{\alpha}).
\end{eqnarray}
The slope of the fitted line in Figure $4.4$ equals the net growth rate $\eta$, which turns out to be approximately $10$\% for the US market when considering the last hundred years, see also Dimson, Marsh and Staunton (2002). The value $\ln(Y_{t})$ is the logarithm of a stationary process at time $t$, which is modeled in (3.3) as a square root process having a gamma density as stationary probability density. This fits well what one observes in Figure $4.4$.  The scaling parameter $\tilde{\alpha}$ can be fitted by ensuring that the average of $Y_{t}$ equals $\frac{a_{1}}{b}$.

Figure $4.5$ shows the normalized S\&P500 which is obtained by dividing the index by the exponential function given in (3.2) using the above estimated values of $\tilde{\alpha}$ and $\eta$ respectively.
Note the stationary density of the square root process in (3.3) has the expression
\begin{eqnarray}
\bar{p}(y)=\frac{(\frac{\delta}{2})^{\frac{\delta}{2}}y^{\frac{\delta}{2}-1}}{\Gamma(\frac{\delta}{2})}\exp\{-\frac{\delta}{2}y\}
\end{eqnarray}
for $y\in (0,\infty),$ where $\delta>2$, see Platen and Health (2006).

\begin{figure}[ht]
\centering
{\includegraphics[width=7.2cm]{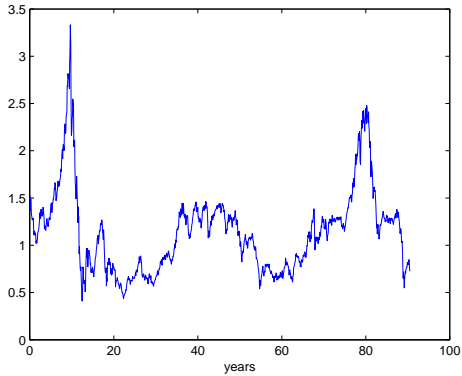}}
\caption{Normalized S\&P $500$ from $1920$ to $2010$.}
\end{figure}

\begin{figure}[ht]
\centering
{\includegraphics[width=7.2cm]{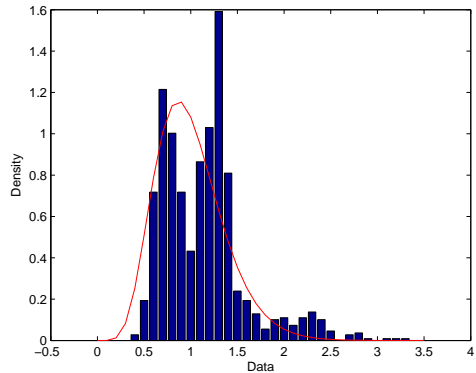}}
\caption{Histogram of the normalized S\&P500 with gamma fit.}
\end{figure}

Figure $4.6$ shows the histogram of the normalized S\&P 500 together with a fit to the above gamma density with estimated dimension $\delta=15$.
Set $b=\eta$ and $a_{1}=b$, then $\gamma=0.0137$.

\section{Risk Neutral $3/2$ Volatility Model}
\renewcommand{\theequation}{5.\arabic{equation}}
\setcounter{equation}{0}
Since the literature on volatility derivatives focus on the risk neutral approach we discuss some important results along these lines of research.
The authors of the two interesting papers:  Carr and Sun (2007) and Itkin and Carr (2009), derived under some assumed risk neutral probability measure a $3/2$ volatility model. It is the generic stochastic volatility model that naturally emerges under the following two plausible assumptions: the \textit{stationary volatility ratio hypothesis} and the \textit{maturity independent diffusion hypothesis}. The first means that the ratio of the volatility of the variance swap rate to the instantaneous volatility of the underlying asset only depends on the variance swap rate and  maturity, but does not depend on calendar time. The second assumption requires that the risk-neutral process for the instantaneous variance is a diffusion process whose coefficients are independent of the variance swap maturity date. For the detailed description of these assumptions and the derivation of the risk neutral $3/2$ volatility model, we  refer to Carr and Sun (2007). This model, which will be described in more detail further below, appears to reflect well empirical properties of short term derivatives on realized variance or volatility. Some of such empirical evidence is described in  Poteshman (1998), Chacko and Viceira (1999), Ishida and Engle (2002), Jones (2003), Javaheri (2004), Bakshi et al. (2004) and Platen and Rendek (2008).

 Because of the strong empirical support that $3/2$ type volatility models attract, this paper places this type of  model in a broader setting, which allows to raise interesting questions from a wider perspective. In particular, the  dynamics of some  $3/2$ models is studied under the real world probability measure $P$. For instance, it links the $3/2$ volatility model to the stylized minimal market model (MMM), described in Platen (2001) and Platen and Heath (2006). The stylized MMM represents a $3/2$ type volatility model when formulated under some assumed risk neutral probability measure, as will be shown below.

 As before, denote by $F_t$ the discounted price at time $t\in[0,T], T<\infty,$ of the underlying diversified equity index. The instantaneous variance of this index shall be denoted by $v_{t}$, which represents its squared volatility.

To remain close to the work by Carr and his co-authors, this section models the stochastic processes under some assumed risk neutral probability measure $Q$.  Assume, as in  Carr and Sun (2007), that the equity index $F_{t}$, or its futures, satisfies the following stochastic differential equation (SDE)
\begin{eqnarray}
dF_{t}=F_{t}\sqrt{v_t}d{\tilde{Z}}_{t},
\end{eqnarray}
 and its squared volatility $v_{t}$ the SDE
\begin{eqnarray}
dv_{t}=\bigg(p(t)v_{t}+q v_{t}^{2}\bigg)dt+\epsilon v_{t}^{3/2}d{\tilde{W}}_{t},
\end{eqnarray}
$t\in[0,T]$. Here $p(t)$, $t\in [0,T]$, and $q$ are assumed to be given real valued, deterministic quantities.
 Furthermore, ${\tilde{Z}}=\{{\tilde{Z}}_{t}, t\in[0,T]\}$ and ${\tilde{W}}=\{{\tilde{W}}_{t}, t\in[0,T]\}$ represent possibly dependent $Q$-standard Wiener processes. The dependence between ${\tilde{Z}}$ and ${\tilde{W}}$ is modeled by the covariation
\begin{eqnarray}
 [{\tilde{Z}},{\tilde{W}}]_{t}=\rho t,
\end{eqnarray}
for $t\in[0,T]$, with given constant correlation parameter $\rho\in[-1,1]$. Note the power $3/2$ for the squared volatility $v_{t}$ in the diffusion coefficient of the SDE (3.7) from which the model derives its name.

 It is well-known that the movements of the squared volatility of a diversified equity index are in reality negatively correlated to those of the index itself. This stylized empirical fact has been called leverage effect, see Black (1976). Realistic calibrations identify a significantly negative correlation parameter $\rho$. Therefore, before going into the analysis of a two-factor model, as studied in Carr and Sun (2007), this paper focuses on the key properties of a one-factor version. To achieve this, the paper sets $\rho=-1$. Thus, a single Wiener process is driving the index, as well as, its volatility. The risk neutral model leaves some ambiguity about the real world dynamics of the diversified index and its volatility. To fix this one can consider the one-factor model that
follows from (5.1) by setting
 \begin{eqnarray}
v_t=\frac{2\gamma\tilde{\alpha}\exp\{\eta t\}}{F_{t}}.
\end{eqnarray}
Here $\tilde{\alpha} >0$ is a scaling parameter, and $\eta>0$ plays the role of the long term average growth rate of the index $F_{t}$ under the real world probability measure $P$. As mentioned earlier, the model reflects in a simple manner the empirical fact that the volatility increases when the index decreases.

The relation (5.4) yields, by (5.1), the SDE
\begin{eqnarray}
dF_t&=& F_t\frac{\sqrt{2\gamma\tilde{\alpha}} \exp\{\frac{\eta}{2} t\}}{\sqrt{F_{t}}}d{\tilde{Z}}_{t}\nonumber\\
&=&\sqrt{2\gamma\tilde{\alpha}} \exp\{\frac{\eta}{2} t\}\sqrt{F_{t}}d{\tilde{Z}}_{t}
\end{eqnarray}
for $t\ge 0$. Under the assumed risk neutral measure $Q$
this is the SDE of a time transformed squared Bessel process of dimension zero, see Revuz and Yor (1999).
 One confirms by application of the It\^{o} formula to (5.4) that by using (5.5) one obtains for $v_{t}$ a $3/2$ volatility model with SDE
\begin{eqnarray}
dv_{t}&=& d\bigg(2\gamma\tilde{\alpha}\frac{\exp\{\eta t\}}{F_{t}}\bigg)\nonumber\\
&=& -\frac{2\gamma\tilde{\alpha}\exp\{\eta t\}}{F_{t}^{2}}\bigg(F_{t}\sqrt{v_t}d{\tilde{Z}}_{t}\bigg)+\frac{2\gamma\tilde{\alpha}\exp\{\eta t\}}{F_{t}}v_tdt
+ \eta\frac{2\gamma\tilde{\alpha}\exp\{\eta t\}}{F_{t}}dt\nonumber\\
&=&v_t\bigg(\eta+v_{t} \bigg)dt-v_{t}^{\frac{3}{2}}d{\tilde{Z}}_{t}
\end{eqnarray}
for $t\in[0,T]$ with $v_{t}>0$. Consequently, the squared
volatility satisfies the SDE (5.2) with the parameter choice $\rho=-1,$ $p(t)=\eta$, $q=1$ and $\epsilon=1$.
Of course, there exist other ways of assuming the real world dynamics of $v_{t}$ other than assuming the functional relationship (5.4). However, such relationship appears to be reasonably plausible and rather tractable.

The squared volatility process, characterized by the SDE (5.6),  has a nonlinear drift and a nonlinear diffusion coefficient. At a   first glance, this seems to make it difficult to obtain statements about its probabilistic properties. However, due to the fact that $F_{t}$ in (5.5) is a squared Bessel process, which is an extremely well studied stochastic process, one knows the analytic form of the probability density of $v_{t}$ and many other properties. As in Carr and Sun (2007)
one may now study the expression
\begin{eqnarray}
 R_t=\frac{2\gamma\tilde{\alpha}}{v_t},
\end{eqnarray}
which equals in our case the value $R_{t}$ of a square root process with
\begin{eqnarray*}
R_{t}=F_{t}\exp\{-\eta t\}.
\end{eqnarray*}
More precisely, one obtains by the It\^{o} formula the SDE
\begin{eqnarray}
dR_t&=&-\frac{2\gamma\tilde{\alpha}}{v_t^{2}}\bigg[v_t\bigg(\eta+v_{t} \bigg)dt-v_{t}^{\frac{3}{2}}d{\tilde{Z}}_{t}\bigg]+\frac{2\gamma\tilde{\alpha}}{v_t^{3}}v_t^{3}dt\nonumber\\
&=& (-\frac{2\gamma\tilde{\alpha}\eta}{v_t}-2\gamma\tilde{\alpha}+2\gamma\tilde{\alpha})dt+2\gamma\tilde{\alpha} v_{t}^{-1/2}d{\tilde{Z}}_{t}\nonumber\\
&=&-\eta R_t dt+\sqrt{2\gamma\tilde{\alpha} R_{t}}d{\tilde{Z}}_{t}.
\end{eqnarray}
It is no surprise that under the assumed risk neutral probability measure $Q$, not only the squared Bessel process $F$ but also the square root process $R=\{R_{t},t\in[0, \infty)\}$ is of dimension $\delta=0$, see Revuz and Yor (1999). Furthermore, according to the just mentioned reference, this process hits the level zero with strictly positive $Q$-probability in any nonzero finite time period. In such an event the volatility explodes, see (5.7). This gives an indication that under $Q$ the expectation of the squared volatility $v_{T}=\frac{2\gamma\tilde{\alpha}}{R_{T}}$ may potentially become infinite. Indeed, under the assumed risk neutral probability measure, the first negative moment of the squared Bessel process $R$ of dimension $\delta=0$ is infinite, see Revuz and Yor (1999).  Consequently, the risk neutral expectation of squared volatility has under the above risk neutral $3/2$ volatility model an infinite value, that is,
\begin{eqnarray}
E^{Q}(v_{T})=\infty.
\end{eqnarray}
 Thus, the typical building block of the payoff of a variance swap on a diversified equity index does not appear to have a finite risk neutral price under the above risk neutral $3/2$ volatility model.

At this point it is important to mention that the assumptions of (5.6) in relation to the parameter choice for the risk neutral $3/2$ volatility model violate those imposed in Carr and Sun (2007) and Itkin and Carr (2009). In particular, these authors require in the SDE (5.2) that the relation $q<\frac{\epsilon^2}{2}$ holds.  As a consequence, the dimension $\delta$ of their square root process $R$ under the assumed risk neutral measure $Q$ is greater than two. This condition guarantees that the volatility will not explode in finite time under the assumed risk neutral probability measure $Q$.  The current paper considers a different model that is not covered by Carr with his coauthors. It is argued that there is no need to restrict the volatility dynamics such that it avoids a volatility explosion under some assumed risk neutral probability measure.
What really matters are the dynamics of the volatility process under the real world probability measure $P$.  Under this probability measure the volatility should be realistically modeled and, thus, not explode, as one observes in reality.
As mentioned in the introduction, also other volatility models, in particular those that model the leverage effect, may create similar  volatility explosions under an assumed risk neutral probability measure resulting in infinite prices, see Andersen and Piterbarg (2007), Lion and Musiela (2007) and Glasserman and Kim (2010).

It has been shown in Platen and Heath (2006), and it is also discussed in Section $7$, that the formally obtained risk neutral price of a nonnegative contingent claim can be significantly higher than the real world price
 to be identified in Section $7$. This means, if there would be a volatility explosion under an assumed risk neutral measure there may still be no volatility explosion under the real world probability measure. Thus, there is a chance that the real world price of  squared volatility could be finite for a $3/2$ volatility model.

\section{Real World $3/2$ Volatility Model}\label{sec.3}
\renewcommand{\theequation}{6.\arabic{equation}}
\setcounter{equation}{0}
To study the dynamics of the underlying diversified equity index and its volatility under the real world probability measure, this paper adopts the benchmark approach described in Platen and Heath (2006). This approach does not require the assumption on the existence of an equivalent risk neutral probability measure. It is only assuming the existence of the num\'{e}raire portfolio for the given market, see Long (1990), Becherer (2001), Karatzas and Kardaras (2007) and Kardaras and Platen (2008).  If there exists an equivalent risk neutral probability measure for a given model, then the benchmark approach recovers fully the results of the risk neutral approach. Otherwise, it still provides a sound derivative pricing methodology, as will be explained in Section $7$.

Under the real world probability measure $P$ one needs also to model the drift in the SDE for the real world dynamics of the diversified index $F_{t}$. This is not necessary under the risk neutral approach and long term trends do not play any role under the classical methodology.

 By (3.4), (3.5) and (3.6) the SDE for $F_t$ under the real world probability measure $P$ has the form:
 \begin{eqnarray}
dF_{t}=F_{t}\bigg((\eta-b+\frac{a_{1}v_t}{2\gamma})dt+\sqrt{v_t}d{Z}_{t}\bigg).
\end{eqnarray}
Here \begin{eqnarray}
Z_t={\tilde{Z}}_{t}-\int_{0}^{t}\theta_{s} ds,
\end{eqnarray}
forms a Wiener process under the real world probability measure $P$, and $\theta_{t}$, see (3.8), is the market price of risk, which provides the link to the process ${\tilde{Z}}$. The
squared volatility for the above $3/2$ volatility model satisfies by (3.7) and (6.2) the SDE
\begin{eqnarray}
dv_{t}=v_{t}\bigg(b +(1-\frac{a_{1}}{2\gamma})v_{t}\bigg)dt-v_{t}^{3/2}d{Z}_{t}
\end{eqnarray}
for $t\in[0,T]$, where $Z$ is the Wiener process as given in (6.2) under $P$.

 The real world dynamics of the process $Y=\{Y_{t}=\frac{2\gamma}{v_{t}},t\ge 0\}$ is given by
\begin{eqnarray}
Y_{t}=\frac{2\gamma}{v_{t}}=\frac{R_{t}}{\tilde{\alpha}}=\frac{F_{t}}{\tilde{\alpha} \exp\{\eta t\}},
\end{eqnarray}
and satisfies the SDE (3.3).
Obviously, the square root process $Y$ has a stationary density. Furthermore, it is known that it never hits zero. Therefore, the above $3/2$ volatility model has no volatility explosion caused by $Y_{t}$ hitting zero. Additionally, the parameter $\eta$ can be interpreted as the long term growth rate of the discounted diversified index, which is a key macro-economic variable.

To elaborate on the interpretation of the real world dynamics of the diversified index one notes from (3.1), (3.2) and (3.3) that the index $F_{t}$ satisfies the SDE
\begin{eqnarray}
dF_{t}=\bigg((\eta-b)F_{t}+a_{1}\tilde{\alpha}\exp\{\eta t\}\bigg)dt+\sqrt{2\gamma\tilde{\alpha}\exp\{\eta t\}F_{t}}dZ_{t},
\end{eqnarray}
where $Z$ is a $P$-Wiener process. Essentially, the deterministic drift in (6.7) models the increase per unit of time in the underlying ``fundamental'' value of the equity index $F_{t}$. On the other hand, the remaining martingale term in (6.7) reflects  the speculative fluctuations of the index. This parsimonious model makes good economic sense, in particular, in the very long term.

Under the real world probability measure the above model appears to be a reasonable model for the dynamics of a well diversified equity index, e.g. the discounted S\&P $500$ total return index. Carr and Sun (2007) and Itkin and Carr (2009) use arguments from variance swap modeling to derive in a plausible manner a risk neutral $3/2$ volatility model.

Recall that under the real world probability measure $P$ the above square root process $Y=\{Y_{t}, t\in [0,T] \}$ has dimension
$\delta=\frac{2a_{1}}{\gamma}>2$, a process with stationary density, never reaching zero under $P$. Under an assumed risk neutral probability measure $Q$ the dimension $\delta$ of $R$, and thus the dimension of the square root process $Y,$ is zero. Consequently, there is a strictly positive risk neutral probability for the event that this process hits zero in a finite time period, see Revuz and Yor (1999). As a result, the measures $Q$ and $P$ do not have the same events of measure zero.
This leads to the conclusion that under the above version of the $3/2$ volatility model the putative risk neutral measure $Q$ is not equivalent to the real world probability measure $P$.

In a complete market, when assuming that $F_{t}$ is the num\'{e}raire portfolio, one has for the corresponding Radon-Nikodym derivative $\Lambda^{Q}_{t}$ at time $t$ the expression
\begin{eqnarray}
\Lambda^{Q}_{t}=\bigg(\frac{F_{t}}{F_{0}}\bigg)^{-1},
\end{eqnarray}
which forms under the above $3/2$ volatility model the inverse of a squared Bessel process of dimension $\delta=\frac{2a_{1}}{\gamma}>2$ under $P$. It is well-known that for $\delta=4$ this process is a local martingale but not a
true $(\underline{\cal{A}},P)$-martingale, see Revuz and Yor (1999). More precisely, $\Lambda^{Q}$ is in this case a non-negative strict $(\underline{\cal{A}},P)$-local martingale, and thus, a strict
$(\underline{\cal{A}},P)$-supermartingale. The above $3/2$ volatility model is then an example of a viable
financial market model, as discussed in Loewenstein and Willard (2000), where the traditional notion of no-arbitrage, see Delbaen and Schachermayer (1998), cannot be verified. Therefore, a more general pricing method than the classical risk neutral one is needed to price derivatives.
\section{Real World Pricing}
\renewcommand{\theequation}{7.\arabic{equation}}
\setcounter{equation}{0}
  It is now the aim to price derivatives under the real world probability measure.  Since an equivalent risk neutral probability measure does not exist under the above model, one can
follow the ideas in Platen and Heath (2006) and use the num\'{e}raire portfolio as num\'{e}raire or benchmark. For this purpose, assume that the diversified index $F_{t}$ represents the num\'{e}raire portfolio and the savings account is the only other traded security to simplify matters. The num\'{e}raire portfolio is defined as the portfolio that when used as benchmark makes all nonnegative benchmarked portfolios supper martingales. This means, the benchmarked savings account is in our setting the inverse of $F_{t}$, and one has by (6.1)
\begin{eqnarray}
d\frac{1}{F_{t}}=\frac{1}{F_{t}}\bigg(b-\eta+v_{t}(1-\frac{a_{1}}{2\gamma})\bigg)dt-\frac{\sqrt{v_{t}}}{F_{t}}dZ_{t}.
\end{eqnarray}
It is well-known that as long as the drift in (7.1) is not strictly positive $\frac{1}{F_{t}}$ forms a super-martingale. Therefore, let us request that
\begin{eqnarray}
\eta\ge b
\end{eqnarray}
and
\begin{eqnarray}
a_{1}\ge 2\gamma.
\end{eqnarray}
We will see below that these parameter constraints are very realistic.

  In this context the following notion turns out to be crucial:\\
{\bf{Definition 7.1}}{\textit{ A price process $U=\{U_{t}, t\in [0,T]\},$ with $E\bigg[ \frac{|U_{T}|}{F_{T}} \bigg]<\infty$, is called \textit{fair} if the corresponding benchmarked price process $\hat{U}=\{{\hat{U}}_{t}=\frac{U_{t}}{F_{t}}, t\in [0,T]\}$ forms an $(\underline{\cal A}, P)$-martingale, that is,
\begin{eqnarray}
{\hat{U}}_t=E\bigg[{\hat{U}}_{\bar{T}}|{\cal A}_t\bigg]
\end{eqnarray}
for all $0\le t\le \bar{T}\le T.$}}

As discussed in Platen and Heath (2006), the minimal supermartingale, which replicates a given benchmarked contingent claim, is the corresponding martingale. Since the minimal price is economically the reasonable price for a replicable claim, it is the fair price that should determine the value of a derivative if no other constraints exist.  For a replicable contingent claim $H_{\bar{T}}$, payable at time $\bar{T}\in [0,T]$ with $E(\frac{|H_{\bar{T}}|}{F_{\bar{T}}})< \infty$, this yields the \textit{real world pricing formula}
\begin{eqnarray}
V_t&=&F_{t}E\bigg(\frac{H_{\bar{T}}}{F_{\bar{T}}}|{\cal A}_t\bigg)
\end{eqnarray}
for all $t\in[0,\bar{T}], \bar{T}\in [0,T].$

Now, one can discuss the link between real world pricing and classical risk neutral pricing.
As shown in Platen and Heath (2006), and as already indicated in (6.6), in a complete market when the Radon-Nikodym derivative process
$\Lambda^{Q}=\{\Lambda_t^{Q}, t\in[0,T]\}$ for the putative risk-neutral probability measure $Q$ is given by the ratio
\begin{eqnarray}
 \Lambda_t^{Q}=\frac{dQ}{dP}|_{{\cal A}_t}=\frac{F_{0}}{F_{t}},
\end{eqnarray}
then, one obtains from the real world pricing formula (7.4) the equivalent expression
\begin{eqnarray}
V_t=E\bigg(\frac{\Lambda_{\bar{T}}^{Q}}{\Lambda_t^{Q}} H_{\bar{T}}|{\cal A}_t\bigg)
\end{eqnarray}
for all $t\in[0,\bar{T}], \bar{T}\in [0,T].$ Recall that zero interest rates are assumed, for simplicity.

In the described special situation of a complete market the Radon-Nikodym derivative process $\Lambda^{Q}$ equals the normalized benchmarked savings account.
If the savings account would be a fair price process, that is, a martingale, then the candidate Radon-Nikodym derivative process $\Lambda^{Q}$ would be an $(\underline{{\cal A}},P)$-martingale. This would guarantee that the risk-neutral probability measure $Q$ exists.  In this case, one would obtain from (7.6) by the Bayes rule the standard risk neutral pricing formula
with $V_t$ equal to $E^Q( H_{\bar{T}}|{\cal A}_t)$.
However, when looking at discounted S\&P $500$ total return index data, one observes that $\Lambda^{Q}_{t}$ exhibits in the long run much smaller values than at the beginning. This is an obvious reflection of the existence of the equity premium. It suggests that it may not be realistic to model $\Lambda^{Q}$  as an $(\underline{{\cal A}},P)$-martingale when pricing derivatives over long periods of time.

Note that in the above $3/2$ volatility model with $F_{t}$ as discounted num\'{e}raire portfolio, the Radon-Nikodym derivative is, in general, not a martingale because it is only a supermartingale. Moreover, it follows from (7.6) by the supermartingale property of the Radon-Nikodym derivative that
 \begin{eqnarray}
 \Lambda_{t}^{Q}\ge E_{t}(\Lambda_{\bar{T}}^{Q})
 \end{eqnarray}
for $0\leq t\leq \bar{T}\leq T$.  For a nonnegative contingent claim $H_{\bar{T}}$, when re-expressing
 (7.7) by using (7.8) one obtains the inequality
\begin{eqnarray}
V_t&\leq& \frac{E\bigg(\frac{\Lambda_{\bar{T}}^{Q}}{\Lambda_t^{Q}} H_{\bar{T}}|{\cal A}_t\bigg)}{E\bigg(\frac{\Lambda_{\bar{T}}^{Q}}{\Lambda_t^{Q}}|{\cal A}_t\bigg)}.
\end{eqnarray}
The right hand side of the above inequality could be interpreted as the formal `` risk neutral'' price, which can be substantially greater than the real world price.
The concept of real-world pricing generalizes classical risk neutral pricing. It does not impose the restrictive condition that $\Lambda^{Q}$ has to form an $(\underline{{\cal A}},P)$-martingale. As a consequence, real world pricing removes from the assumptions of Carr and Sun (2007) and Itkin and Carr (2009) the necessity to require the condition $q<\frac{\epsilon^2}{2}$, which is imposed by these authors to prevent the risk neutral volatility from exploding. The gained freedom allows us to focus on the modeling of the real world dynamics of the $3/2$ volatility model.

\section{Analytical Formulas of Variance Swaps}\label{sec.5}
\renewcommand{\theequation}{8.\arabic{equation}}
\setcounter{equation}{0}
The evaluation of the real world price $V_{v}(t, F_{t})$ of the variance swap for a discounted diversified equity index $F_{t}$ at time $t=0$ is given by:
\begin{eqnarray}
V_{v}(0, F_{0}) &=& F_{0} E [ \frac{ L ({\tilde{\sigma}}^2_{0,T}- K_v )}{F_{T}}] \nonumber\\
&=& F_{0}L E(\frac{{\tilde{\sigma}}^2_{0,T} }{F_{T}}) -  F_{0}L K_v E(\frac{1}{F_{T}})\ .
\end{eqnarray}
Hence, this valuation of a variance swap can be reduced to
the problem of calculating the expectation of the benchmarked underlying variance
$F_{0}E (\frac{{\tilde{\sigma}}^2_{0,T} }{F_{T}})$ and the fair zero coupon bond price $P_{T}(0,F_{0})=F_{0}E(\frac{1}{F_{T}})$.

Since the contract is equal to zero at inception, the fair strike $K_{v}$, or called the variance swap rate, is given by
\begin{eqnarray}
K_{v}=\frac{E (\frac{{\tilde{\sigma}}^2_{0,T} }{F_{T}})}{E(\frac{1}{F_{T}})}.
\end{eqnarray}

For completeness, we shall present the
transition density function~~\\
$p_{\delta}(\varphi_{t},x_{t};\varphi_{\bar{T}},x_{\bar{T}})$
of a time-transformed squared Bessel process~~\\
$X=\{X_{\varphi_{t}},\,\varphi_{t}\in[\varphi_{0},
\varphi_{\bar{T}}]\}$ of
dimension $\delta>2$, which refers to a move from
$x_{t}=X_{\varphi_{t}}$ at the transformed time $\varphi_{t}$ to the level
$x_{\bar{T}}=X_{\varphi_{\bar{T}}}$ at a later transformed time
$\varphi_{\bar{T}}$. From Revuz and Yor (1999), we have
\begin{equation}
p_{\delta}(\varphi_{t},x_{t};\varphi_{\bar{T}},x
    _{\bar{T}}) = \frac{1}{2(\varphi_{\bar{T}}-\varphi_{t})}
    \,\left(\frac{x_{\bar{T}}}{x_{t}}\right)
    ^{\frac{\bar{\nu}}{2}}\exp\left\{-\frac{x_{t}+x
    _{\bar{T}}}{2(\varphi_{\bar{T}}-\varphi_{t})}\right\}\,
    I_{\bar{\nu}}\left(\frac{\sqrt{x_{t}\,x
    _{\bar{T}}}}{\varphi_{\bar{T}}-\varphi_{t}}\right)
\end{equation}
for $\varphi_{t}\in[\varphi_{0},\varphi_{\bar{T}}],$ where $I_{\nu}$ is the modified Bessel function of the first kind with index $\bar{\nu}=\frac{\delta}{2}-1$.
Then for $\varphi\in(0,\infty)$ and $\delta >2$ one can show that the $\tilde{\beta}$th moment
\begin{eqnarray}   E(X^{\tilde{\beta}}_{\varphi}|{\cal A}_{0}) = \left\{ \begin{array}{lr}(2\varphi)^{\tilde{\beta}}\exp\{\frac{-X_{0}}{2\varphi}\}\sum_{k=0}^{\infty}\left(\frac{X_0}{2\varphi}\right)^{k}\frac{\Gamma(\tilde{\beta}+k+\frac{\delta}{2})}{k!\Gamma(k+\frac{\delta}{2})} & \mbox{for}\quad \tilde{\beta}>-\frac{\delta}{2}\\  \infty & \mbox{for} \quad\tilde{\beta} \leq -\frac{\delta}{2}. \end{array}   \right.
\end{eqnarray}
Note that for $\tilde{\beta}\leq -\frac{\tilde{\beta}}{2}$ the corresponding moment does not exist.
The fair price of zero coupon bond $P_{T}(t,F_{t})$ at  maturity $T$ is
\begin{eqnarray}
P_{T}(t,F_{t})&=& F_{t}E(\frac{1}{F_{T}}|{\cal A}_{t})\nonumber\\
&=& \frac{F_{t}}{A_{T}}E\bigg(\frac{1}{Y_{T}}|{\cal A}_{t}\bigg)\nonumber\\
&=&\frac{F_{t}}{A_{T}}E\bigg(\frac{1}{\exp\{-b T\}Y_{\varphi(T)}}|{\cal A}_{t}\bigg)\nonumber\\
&=&\frac{F_{t}}{\exp\{-b T\}A_{T}}E\bigg(\frac{1}{Y_{\varphi(T)}}|{\cal A}_{t}\bigg)\nonumber\\
&=&\frac{F_{t}}{\exp\{-b T\}A_{T}} (2(\varphi(T)-\varphi(t)))^{-1}\exp\bigg\{\frac{-Y_{\varphi(t)}}{2(\varphi(T)-\varphi(t))}\bigg\}\nonumber\\
&\times&\frac{\Gamma(\frac{a_{1}}{\gamma}-1)}{\Gamma(\frac{a_{1}}{\gamma})}{}_{1}F_{1}\bigg(\frac{a_{1}}{\gamma}-1,\frac{a_{1}}{\gamma} , \frac{Y_{\varphi(t)}}{2(\varphi(T)-\varphi(t))}\bigg),\nonumber\\
\end{eqnarray}
where $\varphi(t)=\frac{{\hat{\alpha}}_{0}}{4 b}(e^{b t}-1)$ and the function ${}_{1}F_{1}(.,.,.)$ is the confluent hypergeometric function defined by
\begin{eqnarray*}
{}_{1}F_{1}(a,b,z)=\sum_{k=0}^{\infty}\frac{\Gamma(a+k)}{\Gamma(a)}\frac{\Gamma(b)}{\Gamma(b+k)}\frac{z^{k}}{k!}.
\end{eqnarray*}
Using (2.1), (3.1) and (3.6)
the mean value of the benchmarked underlying variance can be rewritten as
\begin{eqnarray}
&&E (\frac{{\tilde{\sigma}}^2_{0,T} }{F_{T}})\nonumber\\
&=& \frac{2\gamma e^{-\eta T}}{\tilde{\alpha} T} E \bigg(\frac{\int_{0}^{T}\frac{ds}{Y_s}}{Y_{T}}\bigg).
\end{eqnarray}

{\bf{Proposition 8.1}} Let $Y=\{Y_{t}: t\ge 0\}$ satisfy the SDE
\begin{eqnarray}
dY_{t}=(a_{1}-b Y_{t})dt+\sqrt{2\gamma Y_{t}}dW(t)
\end{eqnarray}
Let $\beta=1+m-\alpha+\nu/2$, $m=\frac{1}{2}(\frac{a_{1}}{\gamma}-1)$, $Y_{0}=y>0$ and $\nu=\frac{1}{\gamma}\sqrt{(a_{1}-\gamma)^{2}+4\mu_{1}\gamma}.$ Then if $m>\alpha-\frac{\nu}{2}-1$,
\begin{eqnarray}
&&E\bigg[\frac{\int_{0}^{T}\frac{ds}{Y_{s}}}{Y_{T}^{\alpha}}\bigg]\nonumber\\
&=&-\frac{d}{d\mu_{1}}\frac{1}{2^{\nu}y^{m}}e^{-\frac{by}{\gamma(e^{bT}-1)}+bmT}\bigg(\frac{be^{bT}}{(e^{bT}-1)\gamma}\bigg)^{-m+\alpha-\frac{\nu}{2}}\nonumber\\
&\times&\bigg(\frac{b^2y}{\gamma^2 \sinh^{2}(\frac{bT}{2})}  \bigg)^{\nu/2}\frac{\Gamma(\beta)}{\Gamma(1+\nu)}{}_1{F_{1}}(\beta,1+\nu,\frac{by}{\gamma(e^{bT}-1)})|_{\mu_{1}=0}.\nonumber\\
\end{eqnarray}
\noindent {\bf{Proof:}}
 Use Corollary $5.8$ of Craddock and Lennox (2009) and observe that the fundamental solution in Corollary $5.8$ reduces to the transition density of a square root process as $\mu_{1}\rightarrow 0$. Therefore, we have
 \begin{eqnarray}
 &&E\bigg(e^{-\mu_{1}\int_{0}^{T}\frac{ds}{Y_{s}}}Y_{T}^{-\alpha}\bigg)\nonumber\\
 &=&\int_{0}^{\infty}z^{-\alpha}p(T,y,z)dz,
 \end{eqnarray}
 where
 \begin{eqnarray}
 p(T,y,z)&=&\frac{b}{2\gamma\sinh(bT/2)}\bigg(\frac{z}{y}\bigg)^{\frac{a_{1}}{2\gamma}-\frac{1}{2}}\nonumber\\
 &\times& \exp\bigg(\frac{b}{2\gamma}\bigg(a_{1} T+(y-z)-\frac{y+z}{\tanh(bT/2)}\bigg)\bigg)\nonumber\\
 &\times& I_{\nu}\bigg( \frac{b\sqrt{yz}}{\gamma\sinh(bT/2)}\bigg).\nonumber\\
 \end{eqnarray}

Then, by using formula 6.643.2 of Gradshteyn and Ryzhik (2000), if $Re(m-\alpha+1+\frac{\nu}{2})>0$, we have
\begin{eqnarray}
&&\int_{0}^{\infty} z^{-\alpha+\frac{a_{1}}{2\gamma}-\frac{1}{2}}\exp\bigg(-\frac{b}{2\gamma}(1+\frac{1}{\tanh(bT/2)}z)\bigg)I_{\nu}\bigg( \frac{b\sqrt{yz}}{\gamma\sinh(bT/2)}\bigg)dz\nonumber\\
&=& \frac{\Gamma(\beta)}{1+\nu}\bigg(\frac{b\sqrt{y}}{2\gamma\sinh(bT/2)}\bigg)^{-1}\exp\bigg(\frac{by}{2\gamma(e^{bT}-1)}\bigg)\nonumber\\
&\times& \bigg(\frac{be^{bT}}{\gamma(e^{bT}-1)}\bigg)^{\alpha-m-\frac{1}{2}}M_{\alpha-m-\frac{1}{2},\frac{\nu}{2}}\bigg(\frac{by}{\gamma(e^{bT}-1)} \bigg),\nonumber\\
\end{eqnarray}
where $M_{k,\mu}(z)$ is the Whittaker functions of the first kind.

Furthermore, from the fact that
\begin{eqnarray}
M_{k,\mu}(z)=e^{-\frac{1}{2}z}z^{\frac{1}{2}+\mu}{}_1{F_{1}}(\frac{1}{2}+\mu-k, 1+2\mu,z),
\end{eqnarray}
which is formula $13.1.32$ of Abramowitz and Stegun (1972).
Hence, along with some calculations, we have
\begin{eqnarray}
&&E\bigg(e^{-\mu_{1}\int_{0}^{T}\frac{ds}{Y_{s}}}Y_{T}^{-\alpha}\bigg)\nonumber\\
&=&\frac{1}{2^{\nu}y^{m}}e^{-\frac{by}{\gamma(e^{bT}-1)}+bmT}\bigg(\frac{be^{bT}}{(e^{bT}-1)\gamma}\bigg)^{-m+\alpha-\frac{\nu}{2}}\nonumber\\
&\times&\bigg(\frac{b^2y}{\gamma^2 \sinh^{2}(\frac{bT}{2})}  \bigg)^{\nu/2}\frac{\Gamma(\beta)}{\Gamma(1+\nu)}{}_1{F_{1}}(\beta,1+\nu,\frac{by}{\gamma(e^{bT}-1)}).\nonumber\\
\end{eqnarray}
The result follows from the fact that
\begin{eqnarray}
E\bigg[\frac{\int_{0}^{T}\frac{ds}{Y_{s}}}{Y_{T}^{\alpha}}\bigg]=-\frac{d}{d\mu_{1}}E\bigg(e^{-\mu_{1}\int_{0}^{T}\frac{ds}{Y_{s}}}Y_{T}^{-\alpha}\bigg)|_{\mu_{1}=0}.
\end{eqnarray}

\noindent $\square$

  To further simplify the result of Proposition $8.1$, the derivatives of the confluent hypergeometric function ${}_1{F_{1}}(a_2,b_2,z)$ with respect to the parameters $a_2$ and $b_2$ are summarized below, see Ancarani and Gasaneo (2008).
 Denoted $G^{1}=G^{1}(a_2,b_2;z)=\frac{d {}_1{F_{1}}(a_2,b_2;z)}{d a_2}$ and $H^{1}=H^{1}(a_2,b_2,z)=\frac{d {}_1{F_{1}}(a_2,b_2,z)}{d b_2}$ the first derivatives of the confluent hypergeometric function ${}_1{F_{1}}(a_2,b_2,z)$ with respect to the parameters $a_2$ and $b_2$.
 \begin{eqnarray}
 G^{1}&=&\frac{z}{b_2}\sum_{m_{1}=0}^{\infty}\frac{(a_2)_{m_{1}}(1)_{m_{1}}}{(b_2+1)_{m_{1}}(2)_{m_{1}}}\frac{z^{m_{1}}}{m_{1}!}\nonumber\\
 &\times&{}_2{F_{2}}(1,a_2+1+m_{1},2+m_{1},b_2+1+m_{1};z)
 \end{eqnarray}
 Similarly,
 \begin{eqnarray}
 H^{1}&=&-\frac{za_2}{b_{2}^{2}}\sum_{m_{1}=0}^{\infty}\frac{(a_2+1)_{m_{1}}(b_2)_{m_{1}}(1)_{m_{1}}}{(b_2+1)_{m_{1}}(b_2+1)_{m_{1}}(2)_{m_{1}}}\frac{z^{m_{1}}}{m_{1}!}\nonumber\\
 &\times&{}_2{F_{2}}(1,a_2+1+m_{1},2+m_{1},b_2+1+m_{1};z)
 \end{eqnarray}
where $(a_2)_{n}=\frac{\Gamma(a_2+n)}{\Gamma(a_2)}$.

The equation (8.8) can be rewritten in form of
\begin{eqnarray}
&&E\bigg[\frac{\int_{0}^{T}\frac{ds}{Y_{s}}}{Y_{T}^{\alpha}}\bigg]\nonumber\\
&=&\ln(2)\frac{1}{2^{\hat{\nu}}y^{m}}h_4 h_3(h_2)^{-m+\alpha-\frac{\hat{\nu}}{2}}(h_1)^{\hat{\nu}/2}h_5\nonumber\\
&+&\frac{1}{2^{\hat{\nu}+1}y^{m}}(h_2)^{-m+\alpha-\frac{\hat{\nu}}{2}}\ln(h_2) h_4 h_3(h_1)^{\hat{\nu}/2}h_5\nonumber\\
&-& \frac{1}{2^{\hat{\nu}+1}y^{m}}(h_1)^{\hat{\nu}/2}\ln(h_1)h_4 h_3(h_2)^{-m+\alpha-\hat{\nu}/2}h_5\nonumber\\
&-&\frac{1}{2^{\hat{\nu}+1}y^{m}}h_3 (h_2)^{-m+\alpha-\hat{\nu}/2}(h_1)^{\hat{\nu}/2}\frac{\Gamma(\hat{\beta})\tilde{\varphi}(\hat{\beta})h_4}{\Gamma(1+\hat{\nu})}\nonumber\\
&\times& {}_1{F_{1}}(\hat{\beta},1+\hat{\nu},\frac{by}{\gamma(e^{bT}-1)})\nonumber\\
&-&\frac{1}{2^{\hat{\nu}+1}y^{m}}h_3 (h_2)^{-m+\alpha-\hat{\nu}/2}(h_1)^{\hat{\nu}/2}\frac{-\Gamma(\hat{\beta})\Gamma(1+\hat{\nu})\tilde{\varphi}(1+\hat{\nu})h_4}{(\Gamma(1+\hat{\nu}))^{2}}\nonumber\\
&\times&{}_1{F_{1}}(\hat{\beta},1+\hat{\nu},\frac{by}{\gamma(e^{bT}-1)})\nonumber\\
&-& \frac{1}{2^{\hat{\nu}}y^{m}}h_3 (h_2)^{-m+\alpha-\hat{\nu}/2}(h_1)^{\hat{\nu}/2}\frac{\Gamma(\hat{\beta})}{\Gamma(1+\hat{\nu})} (\frac{1}{2}G^{1}+H^{1})h_{4},
\end{eqnarray}
where $\tilde{\varphi}$ is a digamma function, $h_1=\frac{b^2y}{\gamma^2 \sinh^{2}(\frac{bT}{2})}$, $h_2=\frac{be^{bT}}{(e^{bT}-1)\gamma}$,  $h_3=e^{-\frac{by}{\gamma(e^{bT}-1)}+bmT}$, $h_4=\frac{2}{\gamma \hat{\nu}}$, $h_5=\frac{\Gamma(\hat{\beta})}{\Gamma(1+\hat{\nu})}{}_1{F_{1}}(\hat{\beta},1+\hat{\nu},\frac{by}{\gamma(e^{bT}-1)})$, $\hat{\nu}=\frac{a_{1}}{\gamma}-1$, $\hat{\beta}=1+\hat{\nu}-\alpha$, $G^{1}$ and $H^{1}$ are given by (8.15) and (8.16) respectively with $a_{2}=\hat{\beta}$, $b_{2}=1+\hat{\nu}$ and $z=\frac{by}{\gamma(e^{bT}-1)}$.

\section{Hedge Ratio}
\renewcommand{\theequation}{9.\arabic{equation}}
\setcounter{equation}{0}
In this section, we shall discuss the hedging of variance swaps.
Different methods on hedging variance swaps, have been proposed
in the literature. These methods include the simple delta hedging,
the delta-gamma hedging, hedging using option portfolios,
hedging using a log contract and the vega hedging, etc.
For a comprehensive overview of various hedging strategies,
see Demeterfi et al. (1999), Howison et al. (2004).
In Section $2$, we show that the log contract does not work for the real world $3/2$ volatility model.
We shall derive the exact hedging formulas in this section. The delta hedging formula is derived in form of
\begin{eqnarray}
&&\frac{d E\bigg[\frac{\int_{0}^{T}\frac{ds}{Y_{s}}}{Y_{T}^{\alpha}}\bigg]}{d y}\nonumber\\
&=& \frac{1}{y^{m+1}}(h_1)^{\hat{\nu}/2}h_3 f_{1}\bigg(-m+\frac{y \hat{\nu}}{2 h_{1}}\frac{b^{2}}{\gamma^2\sinh^{2}(\frac{bT}{2})}-\frac{by}{\gamma(e^{bT}-1)}\bigg)\nonumber\\
&+& \frac{1}{y^{m}}(h_1)^{\hat{\nu}/2}h_3 \frac{\Gamma(\hat{\beta})}{\Gamma(1+\hat{\nu})}\frac{\hat{\beta}}{1+\hat{\nu}}{}_1{F_{1}}(\hat{\beta}+1,2+\hat{\nu},\frac{by}{\gamma(e^{bT}-1)})\nonumber\\
&\times& (h_2)^{-m+\alpha-\frac{\hat{\nu}}{2}}h_{4}\bigg(2\ln(2)+\ln(h_{2})-\ln(h_{1})-\tilde{\varphi}(\hat{\beta})+\tilde{\varphi}(1+\hat{\nu})    \bigg)\nonumber\\
&+& \frac{1}{y^{m}}(h_1)^{\hat{\nu}/2}h_3 \bigg\{-\frac{1}{h_1}\frac{b^2}{\gamma^{2}\sinh^{2}(\frac{bT}{2})}h_{4}(h_2)^{-m+\alpha-\hat{\nu}/2}h_{5}\nonumber\\
&-& 2(h_2)^{-m+\alpha-\hat{\nu}/2}\frac{\Gamma(\hat{\beta})}{\Gamma(1+\hat{\nu})}h_{4}\bigg[\frac{b}{2\gamma (e^{bT}-1)(1+\hat{\nu})}\sum_{m_{1}=0}^{\infty} \frac{(\hat{\beta})_{m_{1}} (1)_{m_{1}}}{(2+\hat{\nu})_{m_{1}}(2)_{m_{1}}}\frac{(\frac{by}{\gamma(e^{bT}-1)})^{m_{1}}}{m_{1}!}\nonumber\\
&\times& {}_2{F_{2}}(1,\hat{\beta}+1+m_{1},2+m_{1},2+\hat{\nu}+m_{1},\frac{by}{\gamma(e^{bT}-1)}) \nonumber\\
 &+&\frac{b^2 y}{2\gamma^{2}(1+\hat{\nu})(e^{bT}-1)^{2}}\sum_{m_{1}=0}^{\infty} \frac{(\hat{\beta})_{m_{1}} (1)_{m_{1}}}{(2+\hat{\nu})_{m_{1}}(2)_{m_{1}}}\frac{m_{1}(\frac{by}{\gamma(e^{bT}-1)})^{m_{1}-1}}{m_{1}!}\nonumber\\
&\times& {}_2{F_{2}}(1,\hat{\beta}+1+m_{1},2+m_{1},2+\hat{\nu}+m_{1},\frac{by}{\gamma(e^{bT}-1)})\nonumber\\
 &+&\frac{by}{2\gamma(1+\hat{\nu})(e^{bT}-1)}\sum_{m_{1}=0}^{\infty} \frac{(\hat{\beta})_{m_{1}} (1)_{m_{1}}}{(2+\hat{\nu})_{m_{1}}(2)_{m_{1}}}\frac{(\frac{by}{\gamma(e^{bT}-1)})^{m_{1}}}{m_{1}!}\frac{\hat{\beta}+1+m_{1}}{(2+m_{1})(2+\hat{\nu}+m_{1})}\nonumber\\
&\times& {}_2{F_{2}}(2,\hat{\beta}+2+m_{1},3+m_{1},3+\hat{\nu}+m_{1},\frac{by}{\gamma(e^{bT}-1)}) \nonumber\\
&-& \frac{b\hat{\beta}}{(1+\hat{\nu})^{2}(e^{bT}-1)\gamma}\sum_{m_{1}=0}^{\infty} \frac{(\hat{\beta}+1)_{m_{1}}(1+\hat{\nu})_{m_{1}}) (1)_{m_{1}}}{(2+\hat{\nu})_{m_{1}}(2+\hat{\nu})_{m_{1}}(2)_{m_{1}}}\frac{(\frac{by}{\gamma(e^{bT}-1)})^{m_{1}}}{m_{1}!}\nonumber\\
&\times& {}_2{F_{2}}(1,\hat{\beta}+1+m_{1},2+m_{1},2+\hat{\nu}+m_{1},\frac{by}{\gamma(e^{bT}-1)}) \nonumber\\
&-& \frac{b^{2}\hat{\beta}y}{(1+\hat{\nu})^{2}(e^{bT}-1)^{2}\gamma^{2}}\sum_{m_{1}=0}^{\infty} \frac{(\hat{\beta}+1)_{m_{1}}(1+\hat{\nu})_{m_{1}} (1)_{m_{1}}}{(2+\hat{\nu})_{m_{1}}(2+\hat{\nu})_{m_{1}}(2)_{m_{1}}}\frac{m_{1}(\frac{by}{\gamma(e^{bT}-1)})^{m_{1}-1}}{m_{1}!}\nonumber\\
&\times& {}_2{F_{2}}(1,\hat{\beta}+1+m_{1},2+m_{1},2+\hat{\nu}+m_{1},\frac{by}{\gamma(e^{bT}-1)}) \nonumber\\
&-& \frac{b\hat{\beta}y}{\gamma(1+\hat{\nu})^{2}(e^{bT}-1)}\sum_{m_{1}=0}^{\infty} \frac{(\hat{\beta}+1)_{m_{1}}(1+\hat{\nu})_{m_{1}} (1)_{m_{1}}}{(2+\hat{\nu})_{m_{1}}(2+\hat{\nu})_{m_{1}}(2)_{m_{1}}}\frac{(\frac{by}{\gamma(e^{bT}-1)})^{m_{1}}}{m_{1}!}\nonumber\\
&\times&\frac{\hat{\beta}+1+m_{1}}{(2+m_{1})(2+\hat{\nu}+m_{1})} {}_2{F_{2}}(2,\hat{\beta}+2+m_{1},3+m_{1},3+\hat{\nu}+m_{1},\frac{by}{\gamma(e^{bT}-1)})
\bigg]   \bigg\},\nonumber\\
\end{eqnarray}
where $\hat{\nu}=\frac{a_{1}}{\gamma}-1$, $\hat{\beta}=1+\hat{\nu}-\alpha$ and
\begin{eqnarray*}
f_{1}&=&2^{-\hat{\nu}-1}\bigg[2\ln(2)h_{4}(h_2)^{-m+\alpha-\hat{\nu}/2}h_{5}+(h_2)^{-m+\alpha-\hat{\nu}/2}\ln(h_{2})h_{4}h_{5}\nonumber\\
&-&\ln(h_{1})h_{4}(h_2)^{-m+\alpha-\hat{\nu}/2}h_{5}
-(h_2)^{-m+\alpha-\hat{\nu}/2}h_{4}h_{5}\tilde{\varphi}(\hat{\beta})+(h_2)^{-m+\alpha-\hat{\nu}/2}h_{4}h_{5}\tilde{\varphi}(1+\hat{\nu}) \nonumber\\
&-& 2(h_2)^{-m+\alpha-\hat{\nu}/2}\frac{\Gamma(\hat{\beta})}{\Gamma(1+\hat{\nu})}(\frac{1}{2}G^{1}+H^{1})h_{4}  \bigg].
\end{eqnarray*}

In addition, the Vega of the variance swap can be calculated by
\begin{eqnarray}
&&\frac{d E\bigg[\frac{\int_{0}^{T}\frac{ds}{Y_{s}}}{Y_{T}^{\alpha}}\bigg]}{d \sigma}=(-\sqrt{2\gamma})y^{\frac{3}{2}}\frac{d E\bigg[\frac{\int_{0}^{T}\frac{ds}{Y_{s}}}{Y_{T}^{\alpha}}\bigg]}{d y},\nonumber\\
\end{eqnarray}
where $\frac{d E\bigg[\frac{\int_{0}^{T}\frac{ds}{Y_{s}}}{Y_{T}^{\alpha}}\bigg]}{d y}$ is given by (9.1).

\section{Numerical Results}
In this section, an example of variance swap rates is shown under the benchmark approach and the risk neutral approach. Recall the values of a parameter set in Section $4$, we have $b=\eta$, $a_{1}=b$ and $\gamma=0.0137$.

Table 1 displays
the prices of the variance swap rate of the benchmark approach and the risk-neutral approach for various
maturities respectively and $y=0.3$.

\begin{center}\begin{minipage}[b]{15cm}
\centerline{\it Table 1: Variance swap rate under the benchmark approach}
\quad\quad\quad\quad{\it and corresponding rates under the risk-neutral approach}%
\begin{scriptsize}\begin{center}\begin{tabular}{|c|c|c|r|r|r|r|r|r|r|r|r|r|r|}
\hline
maturities & rates under the benchmark approach & rates under the risk-neutral approach \\%

\hline%
0.25 & 0.137767 & 0.480415 \\%
\hline
0.5 & 0.18197& 0.434351 \\%
\hline
1 & 0.257293& 0.368659 \\%
\hline
1.5 & 0.31596 & 0.323736\\%
\hline
2 & 0.360989& 0.290785 \\%
\hline

\end{tabular}\end{center}\end{scriptsize}\end{minipage}\end{center}%

\section{Conclusions}
This paper studies a $3/2$ volatility model. It is shown that the risk neutral methodology can not handle this kind of $3/2$ volatility model. However, the benchmark approach works well for this kind of model. Furthermore, the analytical formulas for pricing and hedging variance swaps are derived. In addition, the calibration and numerical example are demonstrated.


\begin{thebibliography}{}

\bibitem{AB}
\newblock Abramowitz, M., \& Stegun,~I. (1972).
\newblock {\em Handbook of mathematical functions, with formulas, graphs, and mathematical tables},
\newblock 10th ed. Dover, New York.



\bibitem{AG}
\newblock Ahn, D., \& Gao,~B. (1999).
\newblock A parametric nonlinear model of term structure dynamics.
\newblock {Review of Financial Studies,\/}~{12}, 721--762.


\bibitem{B}
\newblock Andersen, L.B.G., \& Piterbarg,~V. (2007).
\newblock Moment explosions in stochastic volatility models.
\newblock {Finance and Stochastics\/}~{11}, 29--50.

\bibitem{A}
\newblock Andreasen, J. (2001).
\newblock Credit explosives.
\newblock {working paper\/}, Bank of America.

\bibitem{BP}
\newblock Bakshi, G., N.~Ju \& H.~Yang (2004).
\newblock Estimation of continuous time models with an application to equity volatility.
\newblock {\em Working paper\/}, University of Maryland.

\bibitem{B}
\newblock Becherer, D.(2001).
\newblock The numeraire portfolio for unbounded semimartingale.
\newblock {\em Finance and Stochastics\/}~{\bf 5}(3), 327--341.


\bibitem{B}
\newblock Black, F. (1976).
\newblock Studies in stock price volatility change.
\newblock {\em Proceedings of the 1976 Business Meeting of the Business and Economic Statistic Section, American Statistical Association\/}, 177--181.

\bibitem{B}
\newblock Brenner, M. \& D.~Galai (1993).
\newblock Hedging volatility in foreign currencies.
\newblock {\em Journal of Derivatives\/}(1), 53--59.


\bibitem{BM}
\newblock Brockhaus, O. \& D.~Long (2000).
\newblock Volatility swaps made simple.
\newblock {\em Risk\/}~{\bf 2}(1), 92--95.

\bibitem{BT}
\newblock Carr, P. \& D.~ Madan (1999).
\newblock Introducing the covariance swap.
\newblock {\em Risk\/}, February, 47--51.

\bibitem{BP}
\newblock Carr, P., H.~Geman, D.~Madan \& M.~Yor (2005).
\newblock Pricing options on realized variance.
\newblock {\em Finance and Stochastics\/}~{\bf 9}(4), 453--475.

\bibitem{BP}
\newblock Carr, P. \& R.~Lee (2009).
\newblock Volatility derivatives.
\newblock {\em Annu. Rev. Financ. Econ.\/}~{\bf 1}, 1--21.


\bibitem{BP}
\newblock Carr, P. \& J.~Sun (2007).
\newblock A new approach for option pricing under stochastic volatility.
\newblock {\em Review of Derivatives Research\/}~{\bf 10}, 87--150.

\bibitem{Ch}
\newblock Chacko, G. \& L.~Viceira (1999).
\newblock Spectral GMM estimation of continuous-time processes.
\newblock {\em Working paper\/}, Harvard University.


\bibitem{C}
\newblock Cont, R. \& P.~Tankov (2004).
\newblock {\em Financial modelling with jump processes},
\newblock Financial Mathematics Series, Chapman\& Hall/CRC.



\bibitem{C}
\newblock Cox, J., J. Ingersoll \& S. ~Ross (1980).
\newblock An analysis of variable rate loan contracts.
\newblock {\em Journal of Finance\/}~{\bf 35}, 389--403.

\bibitem{C}
\newblock Cox, J., J. Ingersoll \& S. ~Ross (1985).
\newblock A theory of the term structure of interest rates.
\newblock {\em Econometrica\/}~{\bf 53}, 385--407.

\bibitem{C}
\newblock Craddock, M. \& K.A. ~Lennox (2009).
\newblock The calculation of expectations for classes of diffusion processes by Lie symmetry methods.
\newblock {\em The Annals of Applied Probability\/}~{\bf 19}, 127--157.


\bibitem{D}
\newblock Delbaen, F. \& W. ~Schachermayer (1998).
\newblock The fundamental theorem of asset pricing for unbounded stochastic processes.
\newblock {\em Math. Ann.\/}~{\bf 312}, 215--250.

\bibitem{D}
\newblock Delbaen, F.\& H.~Shirakawa (1997).
\newblock Squared Bessel processes and their applications to the square root interest rate model.
\newblock {\em Preprint. Department of Industrial Engineering and Management, Tokio Institute of Technology\/}

\bibitem{CA}
\newblock Demeterfi, K., E.~Derman, M.~Kamal \& J.~Zou (1999).
\newblock A guide to volatility and variance swaps.
\newblock {\em The Journal of Derivatives\/}~{\bf 6}(4), 9--32.

\bibitem{D}
\newblock Dimson, E., P.~Marsh \& M.~Staunton (2002).
\newblock {\em Triumph of the Optimists: $101$ Years of Global Investment Returns},
\newblock Princeton University Press.


\bibitem{D}
\newblock Dupire, B. (1993).
\newblock Model art.
\newblock {\em Risk\/}, Sept: 118--120.


\bibitem{CA}
\newblock Elliott, R. J., T.K.~Siu \& L.~Chan (2007).
\newblock Pricing volatility swaps under Heston's stochastic volatility model with regime switching.
\newblock {\em Applied Mathematical Finance\/}~{\bf 14}(1), 41--62.

\bibitem{RY}
\newblock Gatheral, J. (2006).
\newblock {\em The volatility surface: A practitioner's guide},
\newblock 1st edn, Wiley.


\bibitem{BP}
\newblock Glasserman, P. \& K.K.~Kim (2010).
\newblock Moment explosions and stationary distributions
in affine diffusion models.
\newblock {\em Mathematical Finance\/}~{\bf 20}, 1--33.

\bibitem{G}
\newblock Gradshteyn, I. S. \& I. M.~Ryzhik  (2000).
\newblock {\em Table of integrals, Series, and products},
\newblock 6th ed. Academic Press, San Diego, CA.



\bibitem{CFP}
\newblock Gr\"{u}nbuchler, A. \& F.~Longstaff  (1996).
\newblock Valuing futures and options on volatility.
\newblock {\em Journal of Banking and Finance\/}~{\bf 20},  985--1001.


\bibitem{H}
\newblock Heston, S. (1999).
\newblock A simple new formula for options with stochastic volatility.
\newblock {\em Technical report\/}, Washington university of St. Louis.

\bibitem{H}
\newblock Howison, S., A.~ Rafailidis \& H.~ Rasmussen (2004).
\newblock On the pricings and hedging of volatility derivatives.
\newblock {\em Applied Mathematical Finance\/}~{\bf 11}(4), 317--346.


\bibitem{I}
\newblock Ishida, I. \& R.~Engle  (2002).
\newblock Modelling variance of variance: The square root, the affine, and the CEV GARCH models.
\newblock {\em Working paper\/}, NYU.



\bibitem{I}
\newblock Itkin, A. \& P.~Carr  (2009).
\newblock Pricing swaps and options on quadratic variation under stochastic time change models-discrete observations case.
\newblock {\em Review of Derivatives Research, In Press \/}

\bibitem{J}
\newblock Javaheri, A., P.~Wilmott \& E.G.~Haug  (2002).
\newblock GARCH and volatility swaps.
\newblock {\em Wilmott Magazine\/}, 1--17.


\bibitem{J}
\newblock Javaheri, A. (2004).
\newblock The volatility process: A study of stock market dynamics via parametric stochastic volatility models and a comparison to the information embedded in option prices.
\newblock {\em Ph.D. dissertation\/}.


\bibitem{Jo}
\newblock Jones, C. (2003).
\newblock The dynamics of stochastic volatility: evidence from underlying and options markets.
\newblock {\em Journal of Econometrics\/}~{\bf 116}, 118--224.

\bibitem{K}
\newblock Karatzas, I. \& C.~Kardaras (2007).
\newblock The numeraire portfolio in semimartingale financial models.
\newblock {\em Finance and Stochastics\/}~{\bf 11}(4), 447--493.

\bibitem{K}
\newblock Kardaras, C. \& E.~Platen (2008).
\newblock On the semimartingale property of discounted asset-price processes.
\newblock {\em Working paper\/}, University of Boston.


\bibitem{K}
\newblock Kelly, J.~R. (1956).
\newblock A new interpretation of information rate.
\newblock {\em Bell Syst. Techn. J.\/}~{\bf 35}, 917--926.


\bibitem{L}
\newblock Lewis, A.L. (2000).
\newblock Option valuation under stochastic volatility.
\newblock {\em Finance Press, Newport Beach.\/}

\bibitem{L}
\newblock Lions, P. \& M.~Musiela (2007).
\newblock Correlations and bounds for stochastic volatility models.
\newblock {\em Annales de l'Institut Henri Poincar´e\/}~{\bf 24}, 1--16.

\bibitem{L}
\newblock Loewenstein M. \& G.A.~Willard (2000).
\newblock Local martingales, arbitrage, and viability: free snacks and cheap thrills.
\newblock {\em Econometric Theory\/}~{\bf 16}(1), 135--161.


\bibitem{L}
\newblock Long, J.B. (1990).
\newblock The numeraire portfolio.
\newblock {\em J. Financial Economics\/}~{\bf 26}, 29--69.

\bibitem{M}
\newblock Matytsin, A. (2000).
\newblock Modeling volatility and volatility derivatives.
\newblock Working paper, Columbia University.



\bibitem{M}
\newblock Markowitz, H. \& N.~Usmen (1996a).
\newblock The likelihood of various stock market return distributions, Part 1: Principles of inference.
\newblock {\em J. Risk and Uncertainty\/}~{\bf 13}, 207--219.


\bibitem{M}
\newblock Markowitz, H. \& N.~Usmen (1996b).
\newblock The likelihood of various stock market return distributions, Part 2: Empirical results.
\newblock {\em J. Risk and Uncertainty\/}~{\bf 13}, 221--247.



\bibitem{N}
\newblock Neuberger, A. (1990).
\newblock Volatility trading.
\newblock {\em Working paper\/}, London Bus. Sch.


\bibitem{P}
\newblock Platen, E. (1997).
\newblock A non-linear stochastic volatility model.
\newblock {\em Technical report\/}, Australian National University, Canberra, Financial Mathematics Research Reports. FMRR 005-97.

\bibitem{P}
\newblock Platen, E. (2001).
\newblock A minimal financial market model.
\newblock In {\em Trends in Mathematics}, 293--301. Birkh\"auser.

\bibitem{P}
\newblock Platen, E. (2005).
\newblock Diversified portfolios with jumps in a benchmark framework.
\newblock {\em Asia-Pacific Financial Markets\/}~{\bf 11} (1), 1--22.


\bibitem{PE5}
\newblock Platen, E. \& D.~Heath (2006).
\newblock {\em A Benchmark Approach to Quantitative Finance},
\newblock Springer Finance. Springer.

\bibitem{PE5}
\newblock Platen, E. \& R.~Rendek (2008).
\newblock Empirical evidence on Student-$t$ log-returns of diversified world stock indices.
\newblock {\em J. of Statistical Theory and Practice\/}~{\bf 2} (2), 233--251.


\bibitem{P}
\newblock Poteshman, A. (1998).
\newblock Estimating a general stochastic variance model from option prices.
\newblock {\em Working paper\/}, University of Chicago.



\bibitem{RY}
\newblock Revuz, D. \& M.~Yor (1999).
\newblock {\em Continuous Martingales and Brownian Motion},
\newblock 3rd edn, Springer.

\bibitem{S}
\newblock Sch\"{u}rger, K. (2002).
\newblock Laplace transforms and suprema of stochastic processes,
\newblock {{\em in} K. Sandmann \& P. Sch\"{o}nbucher (eds), \em Advances in finance and stochastics: essays in honour of Dieter Sondermann}, 285--294.
\newblock Springer.


\bibitem{S}
\newblock Spencer, P. (2003).
\newblock Coupon bond valuation with a non-affine discount yield model.
\newblock {\em working paper}, Department of Economics, University of York.

\bibitem{S}
\newblock Swishchuk, A. (2004).
\newblock Modeling of variance and volatility swaps for
financial markets with stochastic volatilities.
\newblock {\em Wilmott magazine\/}~{\bf 2}, 64--72.
\end{thebibliography}
\end{document}